\documentclass[12pt]{article}
\begin{document}
\def\version{File S03v6b.tex last changed 15 October 2000 by P.Gilkey}
\def\nmonth{\ifcase\month\ \or January\or
   February\or March\or April\or May\or June\or July\or August\or
   September\or October\or November\else December\fi}
\def\nmonth{\ifcase\month\ \or January\or
   February\or March\or April\or May\or June\or July\or August\or
   September\or October\or November\else December\fi}
\def\rightheadline{\hfill\folio\hfill}
\def\leftheadline{\hfill\folio\hfill}
\def\operatorname#1{{\rm#1\,}}
\def\text#1{{\hbox{#1}}}
\newtheorem{theorem}{Theorem}
\newtheorem{conjecture}[theorem]{Conjecture}
\newcommand{\reals}{\mbox{${\rm I\!R }$}}
\newcommand{\nats}{\mbox{${\rm I\!N }$}}
\newcommand{\intgs}{\mbox{${\rm Z\!\!Z }$}}
\newcommand{\complex}{\mbox{${\rm C\!\!\!I}$}}
\def\beq{\begin{eqnarray}}
\def\eeq{\end{eqnarray}}
\newcommand{\nn}{\nonumber}
\def\cupd{\buildrel\ldotp\over\sqcup}
\def\trace{\operatorname{Tr}}
\def\birdy{\textstyle\frac\partial{\partial\varepsilon}|_{\varepsilon=0}}
\font\bigfont=cmbx10 scaled \magstep4
\font\medfont=cmbx10 scaled \magstep2
\centerline{\bigfont On properties of the asymptotic}\medbreak
\centerline{\bigfont expansion of the heat trace for}\medbreak
\centerline{\bigfont the N/D problem}\bigbreak
\centerline{\medfont
J. S. Dowker\footnote{Research Partially supported by the EPSRC under
Grant No GR/M45726},
P. B. Gilkey\footnote{Research partially supported by the NSF (USA)}
${}^{\!\!\!,3}$,
and
K. Kirsten$\phantom{.}{}^{\!\!1,}$\footnote{Research partially supported by the
MPI (Leipzig, Germany)}}
\begin{abstract} The spectral problem where the field satisfies Dirichlet
   conditions on one part of the boundary of the relevant domain and Neumann on
   the remainder is discussed. It is shown that there does not exist a classical
 asymptotic expansion for short time in terms of fractional powers of $t$ with
 locally computable coefficients. MSC Classification: 58G25\end{abstract}
\section{Introduction}

Let $M$ be a compact $m$ dimensional Riemannian manifold with smooth
boundary $\partial M$. Let $D$ be an operator of Laplace type on the space
of smooth sections to a vector bundle $V$. Let $D_{\mathcal{B}}$ be the
realization of $D$ with respect to the boundary conditions defined by a
suitable local boundary operator $\mathcal{B}$; we assume $D_{\mathcal{B}}$
is self adjoint. In this paper, we shall study the heat trace asymptotics.

We begin by reviewing the situation in the classical setting and refer to
\cite{refgreinera, refgrubba, refseeley} for further details. We suppose
given a decomposition of $\partial M=C_N\cup C_D$ as the {\it disjoint}
union of two closed (possibly empty) sets. On $\partial M$, let $u_{;m}$
be the covariant derivative of $u$ with respect to the inward unit normal; we
use the natural connection defined by $D$ -- see \cite{refgilkey} for details.
Let the boundary operator
$$\mathcal{B}u:=u|_{C_D}\oplus (u_{;m}+Su)|_{C_N}$$
define Dirichlet boundary conditions on $C_D$ and Robin boundary conditions on $C_N$. Let $\phi$ be
the initial temperature distribution. The subsequent heat temperature distribution
$u:=e^{-tD_{\mathcal B}}\phi$ is defined by the equations:
$$(\partial_t+D)u=0,\ u(0;x)=\phi(x),\text{ and }\mathcal{B}u=0.$$
Let $\{\phi_i,\lambda_i\}$ be a discrete spectral resolution of
$D_{\mathcal{B}}$; the $\phi_i$ are smooth sections of $V$ which form a complete
orthonormal basis for $L^2(V)$ so that
$$\mathcal{B}\phi_i=0\text{ and }D\phi_i=\lambda_i\phi_i.$$
The fundamental solution of the heat equation is trace class. We define
$$
a(D,\mathcal{B})(t):=\text{Tr}_{L^2}e^{-tD_{\mathcal{B}}}=\textstyle\sum_ie^{-t\lambda_i}.$$
\begin{theorem}\label{arefa} Let $C_N\cap C_D=\emptyset$. The heat trace
$a(D,\mathcal{B})(t)$ has a complete
asymptotic expansion as $t\downarrow0$ of the form:
$$a(D,\mathcal{B})(t)\sim\textstyle\sum_{n\ge0}a_n(D,\mathcal{B})t^{(n-m)/2}.$$
The asymptotic coefficients are locally computable as the integral of smooth local
invariants:
\begin{eqnarray*}
a_n(D,\mathcal{B})&=&\textstyle\int_Ma_n(x,D)dx
   +\int_{C_N}a_n^+(y,D,\mathcal{B})dy\\&&
   +\textstyle\int_{C_D}a_n^-(y,D,\mathcal{B})dy.\end{eqnarray*}
\end{theorem}

These invariants have been computed for $n\le5$, see for example
\cite{refbranson, refbransona, refkennedy, refmoss, refmossa}. There exists
a canonical connection $\nabla$ and a canonical endomorphism $E$ so
that $D=-\text{Tr}(\nabla^2+E)$. If $D$ is the Laplacian on $p$ forms, then
$\nabla$ is the Levi-Civita connection and $E$ is given in terms of the
Riemann curvature tensor $R$ by the Weitzenb\"ock formulas. Let indices
$i,j$ range from $1$ through $m$ and index a local orthonormal frame
$\{e_i\}$ for the tangent bundle of $M$. Near the boundary we choose an
orthonormal frame so
$e_m$ is the inward unit normal; let indices $a,b$ range from $1$ through $m-1$
and index the induced frame for the tangent bundle of the boundary. Let
$L_{ab}^{\partial M}$ be the second fundamental form of $\partial M\subset
M$. We adopt the Einstein convention and sum over repeated indices.
\begin{theorem}\label{arefb}\ \begin{enumerate}
\item $a_0(D,\mathcal{B})=(4\pi)^{-m/2}\int_M\trace(I_V) $.
\item $a_1(D,\mathcal{B})=
(4\pi)^{(1-m)/2}\frac14\big\{\int_{C_N}\trace(I_V)-\int_{C_D}\trace(I_V)\big\}$.
\item
$a_2(D,\mathcal{B})=(4\pi)^{-m/2}\textstyle{1\over6}
   \big\{\int_M\trace(R_{ijji}I_V+6E)
    +\int_{C_N}\trace(2L_{aa}I_V+12S)$
\par\qquad$+\int_{C_D}\trace(2L_{aa}I_V)\big\}$.
\item $a_3(D,\mathcal{B})=-(4\pi)^{(1-m)/2}\frac1{384}\big\{
\int_{C_N}\trace(96E+96SL_{aa}+192S^2$
\par\qquad$+(16R_{ijji}-8R_{amma}+13L_{aa}L_{bb}
+2L_{ab}L_{ab})I_V)$\par\qquad$+
\int_{C_D}\trace(96E+(16R_{ijji}
   -8R_{amma}+7L_{aa}L_{bb}-10L_{ab}L_{ab})I_V)\big\}$.
\end{enumerate}\end{theorem}

In the classical setting, $C_N\cap C_D$ is empty so the Neumann and
Dirichlet components do not overlap. There are, however, physically
reasonable settings where $\Sigma:=C_D\cap C_N$ is a non--empty smooth
submanifold of $\partial M$ of dimension $m-2$. Drop a solid ball at
initial temperature $\phi$ into icewater. Supposing it floats, the part of
the boundary of the ball which is in air satisfies Neumann conditions
and the part underwater satisfies Dirichlet conditions.
Here, $\mathcal{B}$
is defined by complementary spherical caps about the north and south poles
of the ball which intersect in a circle of latitude.

The setting where $\Sigma$ is not empty is known in the literature as the
$N/D$ problem. It has been investigated extensively from the
functional analytic point of view \cite{LM, Peetre, Pryde, Simanca}. However,
there are only some preliminary results \cite{refavramidi1,
refavramidi2, refdowker} available concerning the heat trace asymptotics. It is
natural to conjecture that Theorem
\ref{arefa} can be generalized to this setting by adding an extra integral over
$\Sigma$ of some suitably chosen local invariant. The point of this note is to
indicate that the situation is not quite so simple. More specifically, we will
show that the following conjecture is {\bf false}.
\begin{conjecture}\label{arefc} Let $C_N\cap C_D$ be a smooth hypersurface in
$\partial M$. The heat
trace
$a(D,\mathcal{B})(t)$ has a complete
asymptotic expansion as $t\downarrow0$ of the form:
$$a(D,\mathcal{B})(t)\sim\textstyle\sum_{n\ge0}a_n(D,\mathcal{B})t^{(n-m)/2}.$$
The asymptotic coefficients are locally computable as the integral of smooth local
 invariants:
\begin{eqnarray*}
a_n(D,\mathcal{B})&=&\textstyle\int_Ma_n(x,D)dx
   +\int_{C_N}a_n^+(y,D,\mathcal{B})dy\\&&
   +\textstyle\int_{C_D}a_n^-(y,D,\mathcal{B})dy
   +\textstyle\int_\Sigma a_n^\Sigma(z,D,\mathcal{B})dz.\nonumber\end{eqnarray*}
\end{conjecture}

Here is a brief outline to the paper. We shall suppose that Conjecture
\ref{arefc} holds and argue for a contradiction. In \S\ref{sectiontwo}, we
discuss some of the functorial properties which the invariants $a_n^\Sigma$
would have. In \S\ref{sectionthree}, we use the local index formula in a
specific situation to show Conjecture \ref{arefc} is false at the $a_3$ level.
In \S\ref{sectionfour} we present some results using a perturbation expansion
around an exactly soluble, but restricted, case
which also relate to this question. In \S\ref{sectionfive} we conclude by
suggesting an alternative form that the expansion might take.

\section{Properties of the local invariants}\label{sectiontwo}

We use dimensional analysis to study these invariants; this involves
studying the behavior of the heat trace under rescaling. Let $\Omega$ be the
curvature tensor of the connection determined by $D$. We assign weight
$2$ to the tensors $R$, $\Omega$, and $E$. We assign weight $1$ to the
tensors $S$ and $L$. We increase the weight by $1$ for each explicit
covariant derivative which appears. It then follows that the integrands which
can be used to compute the invariants $(a_n^M,a_n^\pm,a_n^\Sigma)$ are universal
polynomials which are weighted homogeneous of degrees
$(n,n-1,n-2)$. For example, Theorem \ref{arefb} expresses $a_2$ in terms of
an interior integral of $\trace(R_{ijji}I_V+6E)$ and boundary integrals of
$\trace(2L_{aa})$ and $\trace(2L_{aa}+12S)$; these expressions have weights
$(2,1,1)$. Thus we see that
$$a_0^\Sigma=0,\ a_1^\Sigma=0,\text{ and }a_2^\Sigma=c_0\dim(V).$$
Calculations of Avramidi \cite{refavramidi1, refavramidi2} and of Dowker
\cite{refdowker} suggest that
$$c_0=-\textstyle\frac{\pi}4(4\pi)^{-m/2}.$$

We shall work with the $a_3^\Sigma$ coefficient to show Conjecture
\ref{arefc} fails. Let  $L^\Sigma$ be the second fundamental form of
$\Sigma\subset C_N$. Any invariant which is homogeneous of weight 1 can be
expressed linearly in terms of the tensors $\{S,L^{\Sigma},L^{\partial
M}\}$.  Thus, in particular, the geometry of the operator $D$ does not
enter into $a_3^\Sigma$. We choose a local frame field on $\Sigma$ so that
$e_{m-1}$ is the unit normal of $\Sigma\subset C_N$ and so that $e_m$ is
the unit normal of $C_N\subset M$. Thus the structure group is $O(m-2)$.
Let $1\le u\le m-2$. We use H. Weyl's theorem \cite{refweyl} on the
invariants of the orthogonal group to see there exist universal constants
$c_i$ so that
\begin{equation}
    a_3^\Sigma(z,D,\mathcal{B})=\trace\{(c_1L_{uu}^\Sigma+c_2L_{uu}^{\partial M}
    +c_3L_{m-1,m-1}^{\partial M})I_V+c_4S\}.\label{brefa}\end{equation}

Let $M=M_1\times M_2$ where $M_2$ is a closed manifold. Let $D:=D_1\otimes
1+1\otimes D_2$ where the operators $D_i$ are operators of Laplace type
over $M_i$. We use a suitable boundary condition $\mathcal{B}_1$ for $D_1$
to induce a corresponding boundary condition $\mathcal{B}$ for $D$. The discrete
spectral resolution for $(D,\mathcal{B})$ is given by the tensor product of the
corresponding discrete spectral resolutions for $(D_1,\mathcal{B}_1)$ and
$D_2$. Therefore
\begin{eqnarray}
&&a(D,\mathcal{B})(t)=a(D_1,\mathcal{B}_1)(t)\cdot a(D_2)(t)\text{ so}
\label{brefb}\\
&&a_n(x,D)=\textstyle\sum_{p+q=n}a_p(x_1,D_1)a_q(x_2,D_2),\nonumber\\
&&a_n^{\pm}(y,D,\mathcal{B})=\textstyle\sum_{p+q=n}a_p^{\pm}(y_1,D_1,\mathcal{B}_1)
      a_q(x_2,D_2),\text{ and}\nonumber\\
&&a_n^\Sigma(z,D,\mathcal{B})=\textstyle\sum_{p+q=n}a_p^\Sigma(z_1,D_1,\mathcal{B}_1)
      a_q(x_2,D_2).\nonumber\end{eqnarray}

{\it A priori}, the constants $c_i$ of equation (\ref{brefa}) could depend
on the dimension $m$ but the usual trick of dimension shifting using
equation (\ref{brefb}) and taking product with a circle shows the constants
$c_i$ are dimension free modulo a multiplicative normalizing factor involving
suitable powers of $4\pi$.

We say that an operator $A:C^\infty(V_1)\rightarrow C^\infty(V_2)$ is of
{\it Dirac type} if the associated second order operators $D^1:=A^*A$ and
$D^2:=AA^*$ are of Laplace type. We assume given boundary conditions
$\mathcal{B}^i$ so that $A$ intertwines the spectral resolutions of
$D^1_{\mathcal{B}^1}$ and $D^2_{\mathcal{B}^2}$. We define:
$$\text{index}(A):=\dim\ \ker(D^1_{\mathcal{B}^1})-\dim\ \ker(D^2_{\mathcal{B}^2}).
$$
The cancellation argument of Bott then shows
\begin{eqnarray}
&&a_m(D^1,\mathcal{B}^1)-a_m(D^2,\mathcal{B}^2)=\text{index}(A)\text{ and}
\label{brefc}\\
&&a_n(D^1,\mathcal{B}^1)-a_n(D^2,\mathcal{B}^2)=0\text{ if }n\ne m.\nonumber
\end{eqnarray}

We shall apply this observation in \S\ref{sectionthree} to the de Rham complex
with absolute or
relative boundary conditions. McKean and Singer \cite{refmckean} used equation
(\ref{brefc}) to prove the Gauss-Bonnet theorem if $m=2$; we refer to
\cite{refgilkeyb} for a
discussion of the higher dimensional setting.

\section{An example on the cylinder}\label{sectionthree}

Let $M:=[0,1]\times S^2$ be the cylinder with the standard product
metric. We consider the de Rham complex and let $\Lambda^{e,o}$ be the bundle of
even and odd differential forms. Let $D^{e,o}$ be the associated
Laplacians. Let $x\in[0,1]$ be the normal variable and let $\Theta\in S^2$
be the angular variable. We then have natural decompositions:
\begin{eqnarray}
   &&\Lambda^{e}_M=\Lambda^{e}_S\oplus dx\wedge\Lambda^{o}_S,\
   D^e_M=(-\partial_x^2+\Delta_S^e)\oplus(-\partial_x^2+\Delta_S^o)\nonumber\\
   &&\Lambda^{o}_M=\Lambda^{o}_S\oplus dx\wedge\Lambda^{e}_S,\
   D^o_M=(-\partial_x^2+\Delta_S^o)\oplus(-\partial_x^2+\Delta_S^e).
   \label{crefa}\end{eqnarray}
If
$\phi_i\in C^\infty(\Lambda_S)$ are differential forms on $M$ taking values in
$\Lambda_S$, then
absolute and relative boundary conditions are defined by the operators:
\begin{eqnarray*}
&&\mathcal{B}_a(\phi_1+dx\wedge\phi_2):=\{\partial_x(\phi_1)\oplus\phi_2\}
\text{ and}\nonumber\\
&&\mathcal{B}_r(\phi_1+dx\wedge\phi_2):=\{\partial_x(\phi_2)\oplus\phi_1\}.
\end{eqnarray*}
We have
\begin{eqnarray*}
 &&(d+\delta)_M(\phi_1+dx\wedge\phi_2)\\
 &&\qquad=\{-\partial_x\phi_2+(d+\delta)_S\phi_1\}
 +dx\wedge\{\partial_x\phi_1-
 (d+\delta)_S\phi_2\}.\end{eqnarray*}
Suppose that $\Delta\phi=\lambda\phi$ on $M$. If $\mathcal{B}^a\phi=0$ on
an open subset ${\mathcal{O}}\subset\partial M$, then
\begin{eqnarray*}
&&\partial_x\{-\partial_x\phi_2+(d+\delta)_S\phi_1\}|_{\mathcal{O}}\\
  &&\qquad=(-\Delta_S+\lambda)\{\phi_2|_{\mathcal{O}}\}+
     (d+\delta)_S\{\partial_x\phi_1|_{\mathcal{O}}\}=0\text{ and}\\
&&\{\partial_x\phi_1-(d+\delta)_S\phi_2\}|_{\mathcal{O}}
  =-(d+\delta_S)\{\phi_2|_{\mathcal{O}}\}=0
\end{eqnarray*}
so ${\mathcal{B}}^a\{(d+\delta)\phi\}=0$ on $\mathcal{O}$. Similarly if
$\mathcal{B}^r\phi=0$ on $\mathcal{O}$, then
\begin{eqnarray*}
&&\{-\partial_x\phi_2+(d+\delta)_S\phi_1\}|_{\mathcal{O}}
  =(d+\delta_S)\{\phi_1|_{\mathcal{O}}\}=0\text{ and}\\
&&\partial_x\{\partial_x\phi_1-(d+\delta)_S\phi_2\}|_{\mathcal{O}}\\
  &&\qquad=(\Delta_S-\lambda)\{\phi_1|_{\mathcal{O}}\}-
     (d+\delta)_S\{\partial_x\phi_2|_{\mathcal{O}}\}=0
\end{eqnarray*}
so ${\mathcal{B}}^r\{(d+\delta)\phi\}=0$ on $\mathcal{O}$. Thus
$(d+\delta)$ preserves the eigenforms of the Laplacian with either absolute or
relative boundary conditions.

Let $\mathcal{B}^-$ denote pure Dirichlet and $\mathcal{B}^+$ pure Neumann
boundary conditions. The
structures decouple and we may decompose
\begin{eqnarray*}
  &&(D^e_M,\mathcal{B}_a)=(-\partial_x^2+\Delta^e_S,\mathcal{B}^+)
      \oplus(-\partial_x^2+\Delta^o_S,\mathcal{B}^-)\\
  &&(D^o_M,\mathcal{B}_a)=(-\partial_x^2+\Delta^e_S,\mathcal{B}^-)
     \oplus(-\partial_x^2+\Delta^o_S,\mathcal{B}^+)\\
  &&(D^e_M,\mathcal{B}_r)=(-\partial_x^2+\Delta^e_S,\mathcal{B}^-)
      \oplus(-\partial_x^2+\Delta^o_S,\mathcal{B}^+)\\
  &&(D^o_M,\mathcal{B}_r)=(-\partial_x^2+\Delta^e_S,\mathcal{B}^+)
      \oplus(-\partial_x^2+\Delta^o_S,\mathcal{B}^-).
\end{eqnarray*}
The interior invariants vanish if $n$ is odd. Thus for dimensional reasons,
\begin{equation}a_3(D^e_M)(x,\Theta)=0\text{ and }
  a_3(D^o_M)(x,\Theta)=0.\label{crefd}\end{equation}
We use Theorem
\ref{arefb} (2) to see that
$$a_1^\pm(y,-\partial_x^2,\mathcal{B}^\pm)=\textstyle\pm\frac14\text{ for }y
\in\partial{[0,1]}.$$
The Dirichlet and Neumann boundary conditions for $\Delta$ decouple on $S^2$.
Since the structures on
$M$ are product, we may apply equation (\ref{brefb}) with
$n=3$, $p=1$, and $q=2$ to the decompositions of display (\ref{crefa}) to compute:
\begin{eqnarray*}
&&a_3^{\partial M}(D^e_M,\mathcal{B}_a)(y,\Theta)=
      \textstyle\frac14\{a_2^S(\Theta,\Delta^e_S)-a_2^S(\Theta,\Delta^o_S)\},
      \\
&&a_3^{\partial M}(D^o_M,\mathcal{B}_a)(y,\Theta)=
     \textstyle\frac14\{-a_2^S(\Theta,\Delta^e_S)
     +a_2^S(\Theta,\Delta^o_S)\},\\
&&a_3^{\partial M}(D^e_M,\mathcal{B}_r)(y,\Theta)=
      \textstyle\frac14\{-a_2^S(\Theta,\Delta^e_S)
        +a_2^S(\Theta,\Delta^o_S)\},\\
&&a_3^{\partial M}(D^o_M,\mathcal{B}_r)(y,\Theta)=
      \textstyle\frac14\{a_2^S(\Theta,\Delta^e_S)
      -a_2^S(\Theta,\Delta^o_S)\}
\end{eqnarray*}
for $y\in\partial\{[0,1]\}$. McKean and Singer \cite{refmckean} showed that
$$\textstyle a_2^S(\Theta,\Delta^e_S)-a_2^S(\Theta,\Delta^o_S)=\frac1{2\pi};$$
this also follows from Theorem \ref{arefb} since $E=0$ for $\Delta^e_S$ while
$\trace{E}=-2R_{ijji}$ for $\Delta^o_S$.
Consequently, we compute the index densities:
\begin{eqnarray}
&&\{a_3^{\partial M}(D^e_M,\mathcal{B}_a)-a_3^{\partial M}(D^o_M,\mathcal{B}_a)\}
(y,\Theta)
       =\textstyle\frac1{4\pi}\nonumber\\
&&\{a_3^{\partial M}(D^e_M,\mathcal{B}_r)-a_3^{\partial M}(D^o_M,\mathcal{B}_r)\}
(y,\Theta)
    =-\textstyle\frac1{4\pi}\label{crefe}
\end{eqnarray}
Let
$C_{a,S}\subset S^2$ and  $C_{r,S}\subset S^2$ be complementary spherical caps about the north and
south pole in
$S^2$. Let $C_{a,M}:=\{0,1\}\times C_{a,S}$ and $C_{r,M}:=\{0,1\}\times C_{r,S}$
give a
corresponding decomposition of the boundary of $M$. Let $\mathcal{B}$ be the
boundary condition
$\mathcal{B}_a$ on $C_{a,M}$ and
$\mathcal{B}_r$ on
$C_{r,M}$. The decompositions of display (\ref{crefa}) induces corresponding
decompositions of $\mathcal{B}$ as the sum of two boundary conditions of the form we have been
considering:
\begin{eqnarray*}
\mathcal{B}(\phi_i)&=&\phi_i|_{C_{a,M}}
     \oplus(\partial_x\phi_i)|_{C_{r,M}}\text{ and}\\
\mathcal{B}(dx\wedge\phi_i)&=&(\partial_x\phi_i)|_{C_{a,M}}
   \oplus\phi_i|_{C_{r,M}}\end{eqnarray*}
As the metric is product, $L_{aa}^{\partial M}=0$. Since $S=0$, the only
non-zero term in $a_3^\Sigma$ given in equation (\ref{brefa}) is
$c_1L_{uu}^\Sigma\trace(I)$. This term is sensitive to the normal of
$\Sigma\subset\partial M$. When studying $D_M^e$ or
$D_M^o$ we have D/N boundary conditions on half the bundle and N/D boundary
conditions on the other half the bundle. Thus this term cancels and we have
\begin{equation}
  a_3^\Sigma(D^e_M,\mathcal{B})(x,\Theta)=0\text{ and }
  a_3^\Sigma(D^o_M,\mathcal{B})(x,\Theta)=0.\label{creff}
\end{equation}
We may therefore use equation (\ref{crefd}), equation (\ref{crefe}), 
and equation
(\ref{creff}) to see
$$a_3(D^e_M,\mathcal{B})-a_3(D^o_M,\mathcal{B})=
   \textstyle\frac1{2\pi}\{\text{vol}(C_{a,M})-\text{vol}(C_{r,M})\}.$$
Thus this difference is not an integer if the spherical caps are not hemispheres. On the other hand,
since $d+\delta$ intertwines the eigenvalues of the Laplacian with either
absolute or relative boundary conditions, we can use equation (\ref{brefc}) to
see that
$$a_3(D^e_M,\mathcal{B})-a_3(D^o_M,\mathcal{B})
   =\text{index}(d+\delta,\mathcal{B})\in\intgs.$$
This contradiction shows that Conjecture
\ref{arefc} is false.

\section{Special cases and perturbative expansions}\label{sectionfour}

The special case of the $N/D$ wedge has been considered locally by Avramidi
\cite{refavramidi2}  and globally by Dowker \cite{refdowker}, where the $N/D$
hemisphere problem was also introduced.
In this section we first enlarge on this last example for which the $N/D$
problem can be solved explicitly in terms of known functions, and then
perturb it to give a more general geometry.

Consider the hemisphere placed at $y\geq 0$ such that its boundary is at
$\varphi =0$ and $\varphi =\pi$ with $\theta \in [0,\pi ]$. The
metric in the standard $(\theta, \varphi)$ polar coordinates and the
Laplacian are given by:
$$ds^2 = d\theta ^2 + \sin ^2 \theta d\varphi ^2 \text{ and }
\Delta = -\frac 1 {\sin ^2 \theta } \frac{\partial ^2}{\partial \varphi ^2}
  -\frac 1 {\sin \theta} \frac \partial {\partial \theta} \sin \theta
\frac \partial {\partial \theta}.$$
We are interested in the eigenfunctions $Y(\varphi, \theta)$
which satisfy the $N/D$ boundary condition. To get an idea of the crucial
difference between these conditions and the classical
Dirichlet and Neumann ones, we will provide details for all of them.

The starting point is the separation of variables,
$
Y(\varphi, \theta) = \Phi (\varphi) \Xi (\theta ) ,\nn
$
which leads to the usual differential equations:
\beq
0 &=& \Phi '' (\varphi ) + \mu^2 \Phi (\varphi ) \nn\\
0 &=& \sin \theta (\sin \theta \,\,\Xi ' (\theta ) ) ' +
( \lambda ^2 \sin ^2 \theta -\mu^2 ) \Xi (\theta ) .\nn
\eeq
With the substitution $\xi = \cos \theta$, $\Xi (\theta ) = u (\xi )$,
one finds
\beq
0&=& (1-\xi^2) u'' (\xi) -2\xi u' (\xi) + \left(
\lambda ^2 -\frac{\mu^2} {1-\xi ^2} \right) u (\xi) .\nn
\eeq
The general solution for $\Phi$ is
$
\Phi (\varphi ) = a \sin (\mu \varphi ) + b \cos (\mu \varphi ) , \nn
$
with $a,b \in \complex$. The equation for $u$ is the differential equation of
the associated Legendre functions. Which eigenfunctions survive as being
linearly independent depends on the value of $\mu$ and thus on
the boundary condition. We refer to \cite{Pockels} for further details.
\medbreak\noindent{\bf 4.1 Dirichlet boundary conditions} mean that
$
\Phi (0) = \Phi (\pi ) = 0 \nn
$
and hence $\mu \in \nats$. Independent solutions are
$
P_l ^\mu (\xi ) \mbox{ and } Q_l ^\mu (\xi )
$
with
$
l\geq \mu, \,l\in \nats, \nn
$
and eigenvalues $\lambda^2 = l (l+1)$. Imposing square
integrability on the eigenfunctions shows that the discrete spectral resolution
is given by the functions:
\beq
Y(\varphi, \theta) = {\cal N}_1 \sin (\mu \varphi ) P_l ^\mu (\cos \theta ) ,
\quad \mu \in \nats, \quad l=n+\mu, \quad n\in \nats_0 , \label{direig}
\eeq
with ${\cal N}_1$ a normalization constant.
For $\theta \to 0$ and $\theta \to \pi$, which is the limit to the north
and south poles, we see that
\beq
P_l^\mu (\cos \theta ) \sim (1-\cos ^2 \theta )^{\mu /2} =
(\sin \theta ) ^\mu \nn
\eeq
and so the eigenfunctions
are differentiable at the poles, which of course, must be the case. We refer
to  \cite{Gromes} for further details.

\medbreak\noindent{\bf 4.2 Neumann boundary conditions} mean that
$\Phi' (0 ) =\Phi' (\pi )=0\nn$. This
yields the quantization condition, $\mu \in \nats_0$ and, arguing as before, we
see that the discrete spectral resolution is given by the functions:
\beq
Y(\varphi, \theta) = {\cal N}_2 \cos (\mu \varphi ) P_l ^\mu (\cos \theta ) ,
\quad \mu \in \nats_0, \quad l=n+\mu, \quad n\in \nats_0 . \nn
\eeq
Again, these are differentiable everywhere, including the poles.
\medbreak\noindent{\bf 4.3 N/D boundary conditions} mean that
$\Phi (0) =\Phi ' (\pi ) =0$. Consequently
$
\mu =  n+1/2, \,\, n\in\nats_0 , \nn
$
or, stated differently,
$
\mu = {\bar m}/ 2 , \,\,\bar m = 1,3,5,...\nn
$
Thus the discrete spectral resolution is given by the functions:
\beq\textstyle
Y(\varphi, \theta) = {\cal N}_3 \sin (\mu \varphi ) P_l ^{-\mu} (\cos \theta ) ,
\quad \mu = \frac{\bar m} 2,\quad \bar m
= 1,3,5,..., \quad n\in \nats_0\,,\nn
\eeq
with $l=n+\mu$.
The vital difference is that now $\mu$ is a half-integer for which the
limiting behaviour of the eigenfunctions near the poles, {\it i.e.} the edge,
is
\beq
P_l ^{-\bar m /2} (\cos \theta ) \sim (\sin \theta )^{\bar m /2} , \nn
\eeq
so that the eigenfunctions are not differentiable at the edge. This is the
crucial difference between the $N/D$ problem and the classical problems, and is
responsible for the non-standard small-$t$ behaviour of the heat-trace.

Having given the eigenfunctions for the hemisphere, for which
the boundary extrinsic curvature vanishes, we now
provide a perturbation approach which allows account to be taken of the
influence of an extrinsic curvature at a boundary. Some general developments
are given first and applied to the hemisphere later.

Assume the unperturbed situation
$\Delta \phi_\lambda = \lambda^2 \phi_\lambda $ with $ \phi \big|_{\partial
 {\cal M} } = 0 $
with non-degenerate eigenvalues. Parametrise the boundary by $y$ and take
$s(y)$ to be the geodesic distance from it. We define the perturbed boundary
$\partial {\cal M}_\epsilon$ by the function $\epsilon s(y)$ with $\epsilon$
very small and call the resulting manifold ${\cal M}_\epsilon$. Then, to order
$\epsilon$, the perturbative formulation of the problem reads as follows,
\beq
\Delta \psi _\alpha = \alpha^2 \psi _\alpha, \quad
\psi_\alpha \big|_{\partial {\cal M} _\epsilon} = 0, \nn
\eeq
where the initial ansatz for the eigenfunctions and eigenvalues is given by:
\beq
\psi_\alpha = \phi_\lambda + \epsilon \phi_\alpha ', \quad
\alpha^2 = \lambda^2 + \epsilon \eta _\alpha . \nn
\eeq
The perturbation of the eigenvalues, $\eta_\alpha$, is determined by:
\begin{eqnarray*}
\eta_\alpha &=& \textstyle\int_{\partial {\cal M_\epsilon}} \phi_\alpha '
\partial _n
\phi _\lambda ^* -  \int_{\partial {\cal M_\epsilon}} \phi_\lambda^*
 \partial _n \phi '\end{eqnarray*}
with the exterior normal $\partial_n$; see, for example, 
\cite{MF}. For Dirichlet
conditions one can use the identity $
\phi'|_{\partial {\cal M}_\epsilon} = -\frac 1 \epsilon \phi_\lambda
\big|_{\partial {\cal M}_\epsilon}\nn
$
together with the expansion
\begin{eqnarray}
\phi_\lambda \big|_{\partial {\cal M}_\epsilon}& =& \phi_\lambda
\big|_{\partial {\cal M}} -\epsilon s(y) \partial_n \phi_\lambda
\big|_{\partial {\cal M}} + ...\nonumber
\\&=& -\epsilon s(y) \partial _n \phi_\lambda
\big|_{\partial {\cal M}} + ... \label{tay1}
\text{ to see}\\
\textstyle\eta_\alpha&=&\textstyle -\frac 1 \epsilon
\int_{\partial {\cal M_\epsilon}} \phi_\lambda \partial _n
\phi _\lambda ^* =-  \int_{\partial {\cal M_\epsilon}} \phi_\lambda^*
 \partial _n \phi '\label{11b}
 \int_{\partial {\cal M}} s(y) |\partial _n \phi_\lambda|^2 dy .
\end{eqnarray}
It is important to note that the expansion (\ref{tay1}) is well defined for the
Dirichlet and Neumann eigenfunctions (\ref{direig}).

In the case of degenerate eigenvalues, which is needed here, the situation is
slightly more complicated. Let $j$ index the degeneracy. In this case one can
obtain the secular equation
\begin{eqnarray}
&&\textstyle\sum_j c_{ij} (\eta_\alpha^i \delta_{kj} + B^D_{kj} ) = 0 ,
\label{degdir}\text{ with}\\
&&\textstyle
B_{kj}^D = -\int_{\partial {\cal M}} (\partial_n \phi_\lambda^{k*}) s(y)
(\partial_n \phi_\lambda^j) . \label{relintdir}
\end{eqnarray}

Similarly, when considering Neumann conditions and when perturbing
the shape of the boundary, the equation analogous to (\ref{degdir})
becomes
\begin{eqnarray}&&\textstyle
\sum_j c_{ij} (\eta_\alpha^i \delta_{kj} + B^N_{kj} ) = 0 ,
\text{ with}\nonumber\\&&\textstyle
B_{kj}^N = \int_{\partial {\cal M}}  \phi_\lambda^{k*} s(y)
(\partial_n^2 \phi_\lambda^j) . \label{relintneu}
\end{eqnarray}

Some of these formulae are implicit in the work of Fr\"ohlich,
\cite{Frohlich}, but can be traced back to Rayleigh, \cite{Rayleigh}.
The extensive discussion in Morse and Feshbach, \cite{MF}, contains all that one
needs. A more mathematical treatment is given by
Garabedian and Schiffer, \cite{GS} Chap.V, based on Green's theorem and
 Hadamard's formula.

We apply these developments to the hemisphere. It seems
natural to displace the entire boundary by an amount, say $\epsilon$,
perpendicular to the equator, {\it i.e.} the rim, thus making the new manifold
a cap. Geometry shows that the
extrinsic curvature, $L^{\partial {\cal M}}$, of the perturbed boundary
equals $\epsilon$. Then the relevant integrals for the eigenvalue perturbations
are,
\beq
B_{\mu \mu'}^D &=&\textstyle -2{\cal N}_1^2 \mu \mu' L^{\partial {\cal M}}
\int_0^\pi \frac{d\theta} {\sin^2\theta}\, P_l^\mu (\cos \theta)\, P_l
^{\mu '} (\cos \theta) ,\text{ and} \label{15aa}\\
B_{\mu \mu'}^N &=&\textstyle-2{\cal N}_2^2  \mu'\,^2 L^{\partial {\cal M}}
\int_0^\pi \frac{d\theta} {\sin^2\theta}\, P_l^\mu (\cos \theta)\, P_l
^{\mu '} (\cos \theta) .\nn
\eeq

The behaviour of the Legendre functions for $\theta \to 0,\pi$
shows that these integrals exist and, although an explicit evaluation
is tedious (they are
surprisingly not listed in standard references like \cite{AS,GR}),
in principle this determines the eigenvalues to order $\epsilon$ and
undoubtedly would reproduce the correct leading heat kernel expansion.
We have not pursued this calculational check.

It seems reasonable to apply the same perturbative approach to the
$N/D$ problem. In this case the $N$ and $D$ parts contribute additively to
the secular equation,
\beq
\sum_j c_{ij} (\eta_\alpha^i \delta_{kj} + B^D_{kj} +B^N_{kj} ) = 0 ,
\nn\eeq
where the definitions (\ref{relintdir}) and (\ref{relintneu}) still hold
but involving the eigenfunctions of the $N/D$ problem.
As one soon realizes, all
integrals exist, {\it except} the ones with $\mu = \mu' = -1/2$.
The reason may be found in the use of (\ref{tay1}) which {\it cannot}
be applied for these modes because it leads to divergences at the edges
of the manifold. To avoid the use of (\ref{tay1}) we
revert to (\ref{11b}) and evaluate just {\it these} awkward modes
on the perturbed boundary, $\partial {\cal M}_\epsilon$, directly.

We return to Dirichlet conditions, where
everything is well defined, to illustrate the situation. The
boundary $\partial {\cal M}_\epsilon$ can be parametrised by noting that
along it,
$
y=\epsilon=\sin \theta \sin \varphi   ,\nn
$
{\it i.e.}
$$
\varphi = \sin^{-1} \left(\frac{\epsilon} {\sin\theta} \right), \quad
x\geq 0 ,\nn\text{ and }
\varphi = \pi - \sin^{-1} \left(\frac{\epsilon} {\sin\theta} \right), \quad
x\leq 0.$$
To leading order in $\epsilon$, the geometrical quantities
(normal derivative and volume element) on $\partial {\cal M}_\epsilon$
agree with those on $\partial {\cal M}$ and, up to irrelevant corrections,
one finds
\beq
B_{\mu \mu'}^D &=& -2 {\cal N}_1^ 2 \mu' L^{\partial M} \frac 1 \epsilon
\int_{\epsilon}^{\pi-\epsilon} \frac{d\theta}{\sin\theta}
     P_l^ {\mu} (\cos\theta) P_l^{\mu'} (\cos \theta) \times \nn\\
& &\quad\quad\quad\quad\sin
\left[ \mu \sin^{-1} \frac \epsilon {\sin\theta} \right]
 \cos \left[ \mu' \sin^{-1} \frac \epsilon {\sin\theta} \right] .
\label{15a}
\eeq
To the relevant order, the integration limits can be set to $0$ and $\pi$
and comparison of (\ref{15aa}) with (\ref{15a}) shows that the expansion
with respect to $\epsilon$
of the eigenfunctions evaluated at $\partial {\cal M}_\epsilon$
leads to the occurrence of a factor
$\epsilon/\sin\theta$ to give agreement with the perturbation form.

Turning to the $N/D$ problem, we are therefore led to consider
integrals of the type
\beq\textstyle
\int_\epsilon^{\pi-\epsilon} \frac{d\theta} {\sin^2 \theta} P_l^{-\mu}
(\cos\theta) P_l^{-\mu'} (\cos\theta) \nn
\eeq
which are well defined in the limit $\epsilon \to 0$ for all values of
$\mu,\mu'$ {\it except} $\mu=\mu' = 1/2$. For these ``critical subspaces",
using the explicit Gegenbauer representation \cite{GR},
\beq\textstyle
P_{\nu-1/2}^{-1/2} (\cos\theta) = \sqrt{\frac 2 {\pi \sin\theta}}
   \frac{\sin(\nu \theta)} \nu , \nn
\eeq
the relevant integrals have the form
\beq\textstyle
\int_\epsilon^{\pi -\epsilon} d\theta \frac{\sin ^2 [(n+1) \theta]}
    {\sin ^3 \theta} ,\nn
\eeq
which behaves like $\log \epsilon = \log L^{\partial M}$ as $\epsilon \to 0$.
All other eigenfunctions yield an ${\cal O} (\epsilon)$
term to leading order and will not change this log behaviour.
These remarks suggest the existence of a term $L^{\partial M} \log
L^{\partial M}$ in the heat trace expansion which is an indirect
indication, via dimensional arguments, of the appearance of $\log t$ terms as
well (see \S\, 5).

Thinking about higher order perturbation theory, it is expected that
modes which are differentiable only $(k-1)$ times, $k\in\nats$, lead to
the occurrence of $\epsilon^k \log \epsilon$ with associated $\log t$ terms
in the heat trace expansion.

We stress that, for the perturbed geometry, the extrinsic curvature
$L^{\partial {\cal M}}$ does not vanish at the edge and this is the cause of
the trouble. When the boundary perturbation, $s(y)$, vanishes at the edges, the
perturbation is well defined and we expect the standard trace expansions to hold.

\section{Conclusions}\label{sectionfive}

We have shown that if there is an asymptotic expansion of the heat trace for
the $N/D$ problem,
then it is not as simple as in the standard setting. It is  possible that
$\log t$ terms enter, the generic behaviour for singular situations.
There is one setting where such terms are known to arise. If instead of local
boundary conditions, spectral conditions are imposed, one has a partial
asymptotic expansion of the form,
$$a(D,\mathcal{B})(t)=\sum_{n<m}a_n(D,\mathcal{B})t^{(n-m)/2}+O(t^{-\frac18}).$$
Again the invariants are locally computable and we refer to \cite{refdgk,
refgk} for formulae if $n\le3$.
However, the complete asymptotic expansion involves non-local and
$\log$ terms \cite{refgrsea,
refgrseb}. Thus perhaps Conjecture \ref{arefc} should be replaced by an
asymptotic
expansion of the form
$$a(D,\mathcal{B})(t)\sim\textstyle\sum_{n\ge0}\bigg(\alpha_n(D,\mathcal{B})
\log t+
     a_n(D,\mathcal{B})\bigg)t^{(n-m)/2}$$
where the leading
term $\alpha_n(D,\mathcal{B})=\textstyle\int_\Sigma\alpha_n^
\Sigma(z,D,\mathcal{B})dz$ is locally
computable and where the difference
\begin{eqnarray*}
    a_n(D,\mathcal{B})-\textstyle\int_Ma_n(x,D)dx
   -\textstyle\int_{C_D}
    a_n^-(y,D,\mathcal{B})dy-\textstyle\int_{C_N}a_n^+(y,D,\mathcal{B})dy\end{eqnarray*}
is a non-local invariant determined by
the behavior of $D$ and $\mathcal{B}$ near $\Sigma$.
As we have argued in Section 4, $\log t$ terms may occur to
compensate the $\log L^{\partial M}$ terms present in the perturbed hemisphere
example. Further study of the heat trace asymptotics of the D/N problem seems
indicated.

\medbreak\hrule
\smallbreak\noindent {\bf J. S. Dowker} Department of Theoretical Physics, The University of Manchester,
Manchester England email: dowker@a13.ph.man.ac.uk
\smallbreak\noindent {\bf P. Gilkey} Mathematics Department, University of Oregon, Eugene Or 97403
email: gilkey@darkwing.uoregon.edu
\smallbreak\noindent {\bf K. Kirsten} Department of Theoretical Physics, The University of Manchester,
Manchester England email: kirsten@a13.ph.man.ac.uk
\vfill\noindent
Version \version\smallbreak\noindent printed\ \number \day\ \nmonth\
\number\year
\end{document}